# Data-Driven Distributed Common Operational Picture from Heterogenous Platforms using Multi-Agent Reinforcement Learning

Indranil Sur[1†]    Aswin Raghavan[1†]    Abrar Rahman[1]
[1]SRI, Princeton, NJ, USA.
{indranil.sur, aswin.raghavan, abrar.rahman}
@sri.com

James Z Hare[2]    Daniel Cassenti[2]    Carl Busart[2]
[2]DEVCOM Army Research Laboratory
{james.z.hare.civ, daniel.n.cassenti.civ,
carl.e.busart.civ} @army.mil

## Abstract

The integration of unmanned platforms equipped with advanced sensors promises to enhance situational awareness and mitigate the "fog of war" in military operations. However, managing the vast influx of data from these platforms poses a significant challenge for Command and Control (C2) systems. This study presents a novel multi-agent learning framework to address this challenge. Our method enables autonomous and secure communication between agents and humans, which in turn enables real-time formation of an interpretable Common Operational Picture (COP). Each agent encodes its perceptions and actions into compact vectors, which are then transmitted, received and decoded to form a COP encompassing the current state of all agents (friendly and enemy) on the battlefield. Using Deep Reinforcement Learning (DRL), we jointly train COP models and agent's action selection policies. We demonstrate resilience to degraded conditions such as denied GPS and disrupted communications. Experimental validation is performed in the Starcraft-2 simulation environment to evaluate the precision of the COPs and robustness of policies. We report less than 5% error in COPs and policies resilient to various adversarial conditions. In summary, our contributions include a method for autonomous COP formation, increased resilience through distributed prediction, and joint training of COP models and multi-agent RL policies. This research advances adaptive and resilient C2, facilitating effective control of heterogeneous unmanned platforms.

## 1 INTRODUCTION

The integration of unmanned platforms equipped with advanced sensors holds promise for mitigating the "fog of war" and elevating situational awareness. However, managing and disseminating the influx of data from such platforms poses a substantial challenge to the information processing capabilities of central Command and Control (C2) nodes, particularly given the exponential growth in data volume with increasing platform numbers. The current manual processing methods are ill-suited for future C2 scenarios involving swarms of unmanned platforms. In this study, we present a framework utilizing a multi-agent learning approach to overcome this barrier.

We consider a framework where agents communicate with each other (and with humans) in an autonomous fashion, and such communication functions are trained in a data-driven manner. At each time step, each agent can send/receive a real-valued message vector. The vector is a learned encoding of the agent's perception or field of view (FoV). The vectors are not easily interpretable by adversaries, allowing for secure message transfer.

On the receiver's side, the message must be decoded to recover the sender's perception and action. Furthermore, the information should be *integrated* (aggregated over time) into a Common Operational Picture (COP). Like the encoder, the decoder is also learned in a data-driven manner. In this paper, we simplify the definition of a COP as the current state (position, health, shield, weapon, etc.) of each friendly and enemy agents on the battlefield. We argue that the COP is essential to decision-making agents.

In recent years, AI/ML approaches that are trained end-to-end in a data-driven manner have shown great promise. In the context of a data-driven autonomous COP, one advantage is that no modelling assumptions are made about the noise in the sensors and actuators, the dynamics of the adversary, etc. With sufficient training, our data-driven method will produce highly precise COPs.

However, ML models can be sensitive to deviations from the training data or training scenarios. This contrasts with the DDIL (denied, disrupted, intermittent, and limited impact) environments, which are typically assumed in army C2 scenarios. Our experiments emphasize evaluation of the resilience to increased fog, denied GPS, and disruption of communications (e.g., jamming).

Data-driven end-to-end training of our encoders and decoders is achieved using deep learning of deep neural networks (DNN). One challenge associated with the application of DNNs to COP formation is the lack of human interpretability in communications. Human interpretability is crucial for a human operator to effectively control the swarm. For example, by interpreting the communication, the operator might understand the features used by the swarm for (autonomous) decision-making. Our method is human-machine *interchangeable*, meaning that a human operator can decode the incoming messages and encode their perceptions to communicate with the swarm. The resulting COP enables human directability of the swarm.

In practice, the COP is heavily used in mission execution, e.g., to ensure coordinated movements. We hypothesize that incorporating the COP into autonomous decision-making agents will produce resilient multi-agent policies (e.g., resilience to changes in the enemy). Our experiments compare multi-agent policy learning with and without the COP against multiple state-of-the-art methods and validate the hypothesis.

Next, we summarize our methodology. We first describe our deep learning formulation in which each agent encodes its perceptions and actions into compact vectors and transmits them. The underlying embedding vector space is shared across agents to enable a shared situational understanding. An encoder-decoder is trained per agent to produce local COPs. The local COP should be consistent with agent perceptions and should predict all units' state (incl. position) over the area of operation.

End-to-end training of the COP is performed jointly with agent policies using Deep Reinforcement Learning (DRL) on a diverse set of simulated scenarios, initial force configurations, and adversary actions. The output of training is an encoder-decoder neural network (NN) and a policy NN shared across agents. The training can be configured in several ways: to minimize bandwidth, maximize resilience to disruption, channel noise, packet loss, jamming of GPS, etc. The method can be applied to coordinated information-gathering missions.

Experiments are performed in the Starcraft-2 (SC2) multi-agent environment [1]. The effectiveness of our method is empirically observed in multiple blue-vs.-red scenarios that are modelled in SC2. Specifically, we test and evaluate our method on the challenging and realistic *TigerClaw* scenario (Figure 1) that was developed by the DEVCOM Army Research Lab (ARL) and Army subject-matter experts (SMEs) at the Captain's Career Course, Fort Moore, Georgia, US [2].

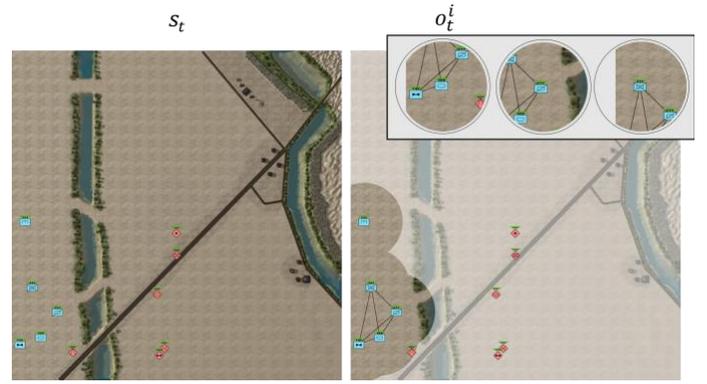

*Figure 1: (Left) Example state from the Tigerclaw scenario. (Right) Each agent's perception (local observation) and communication links between them.*

The COPs are evaluated for accuracy and hallucinations that reveal interesting training dynamics. Our method produces highly accurate COPs with less than 5% error (compared to ground truth) over the entire simulation. To test the robustness of policies, we compare our method to multiple state-of-the-art multi-agent RL methods and baselines. We show that our method produces policies resilient to degraded visual range, degraded communication, denied GPS, and changes in the scenario.

In summary, this research enables the command and control of heterogeneous autonomous platforms with human-in-the-loop through data-driven COP formation and advances the field of adaptive and resilient C2. The contributions are as follows:

- A method to autonomously form an interpretable Common Operational Picture (COP) in real-time, including the prediction of enemy positions over the entire area of operation.
- Demonstrate increased resiliency to denial of visual range and GPS because of the distributed COP prediction using inter-agent communication.
- Increased overall mission success by joint training of COP models and multi-agent RL policies.

## 2 PROBLEM FORMULATION

Consider a multi-agent system consisting of $N$ agents. Each agent is defined as the tuple $(\mathcal{O}^i, \mathfrak{A}^i, \mathcal{C}_{in}, \mathcal{C}_{out}, \mathfrak{P}^i)$, for agent $i = 1 \dots, N$. The agents' perception of the environment is a mapping $\mathcal{O}^i : \mathfrak{S} \times \mathfrak{P}^i \to \mathbb{R}^d$ from the underlying state of the world ($s \in \mathfrak{S}$) and agent-specific capabilities $\mathfrak{P}^i$ (e.g., field-of-view) to a $d$-dimensional observation vector $o^i = \mathcal{O}(s, p_i) \in \mathbb{R}^d$ (dropping the

Observations of three agents:

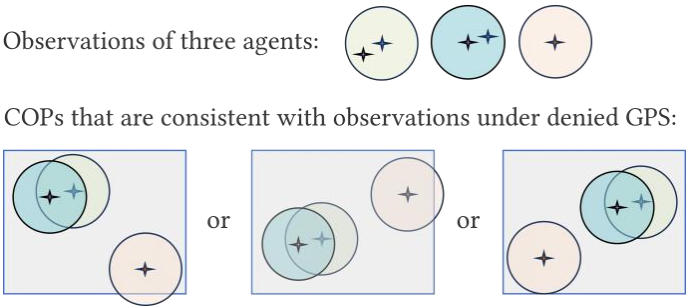

COPs that are consistent with observations under denied GPS:

or ... or ...

*Figure 2: Illustration of the challenge in COP formation in the face of GPS denial. Colors represent different agents.*

dependence on time for brevity). Figure 1 shows an example of state and agent observations. In this paper, agent observations and actions are in the agent's frame of reference i.e., egocentric, whereas states are represented in a global frame of reference. For example, agent observations might include range, bearing, and health of observed units, whereas the state will include ground truth about all quantities of all units (blue-vs-red).

At each time step, each agent receives a $C$-dimensional communication message from each agent from the previous time step. The messages are processed using the function $\mathcal{C}_{in}: \mathbb{R}^{C^N} \to \mathbb{R}^C$. In this paper, denied communication and out-of-range communication is represented as zero-valued vectors at the receiver $\mathcal{C}_{in}$. We do not assume stable communication pathways to be able to send and receive transmissions; rather, messages are "zero-ed" out in the communication channel unbeknownst to the sender.

At each time step, each agent sends a $C$-dimensional communication message using the function $\mathcal{C}_{out}: \mathcal{O}^i \times \mathfrak{A}^i \times \mathbb{R}^C \to \mathbb{R}^C$ that maps the agent's local observation, action at the current time step and the result of $\mathcal{C}_{in}$ to an output message. Information to uniquely identify the agent can also be transmitted by including a unique ID for each agent in the observation space.

The functions $\mathcal{C}_{in}$ and $\mathcal{C}_{out}$ are represented as neural networks (NN) with learnable weights. The NN weights are shared across agents (and human operators).

Each agent can take an action $a^i$ from its action set $\mathfrak{A}^i$ that affects the evolution of the state. The action set can be fixed or a function of capabilities $\mathfrak{P}^i$ or observation $o_i$. We train a policy shared across agents $\pi: \mathcal{O}^i \times \mathbb{R}^{C^N} \to \mathfrak{A}^i$ that maps its local observation and messages from other agents to an action in its own action set.

In one sense, the multi-agent system forms a distributed mobile sensor network where each agent senses only a part of the world. Inference of the underlying state (including friendly and enemy positions) corresponding to the agents' local observations is a key component of situational awareness and is necessary for developing a Common Operational Picture (COP), i.e., a global understanding of the uncertainty in the underlying state as the battle evolves. In the distributed setting, each agent forms a local prediction of the COP via communication and propagation of other agent's local COP, observation and action, which allows for uncertainty reduction and improved situational awareness.

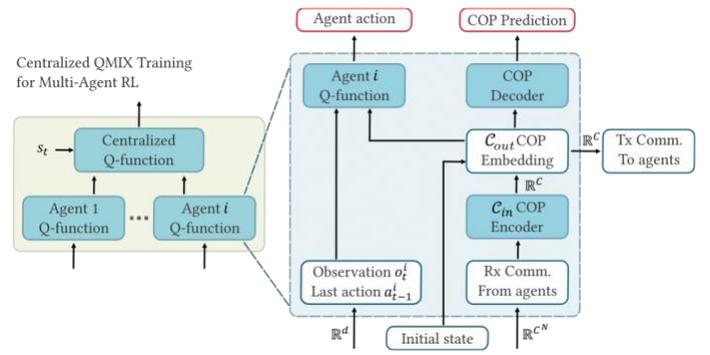

*Figure 3: Overview of our framework for COP prediction from learned communication. The COP is determined and used in the decision-making process. We use QMIX [4] as an example MARL method for COP integration.*

For example, a friendly agent could observe the range and bearing of enemy units in its local frame of reference (e.g., a tank at 1200 meters, 30 degrees north). Given this information, a different friendly agent in another part of the battlefield needs to infer the tank's relative position to itself. This is challenging, especially when global positioning is denied. This challenge is illustrated in Figure 2. Our COP enables a solution to this challenge using the shared embedding space for communications.

Figure 3 shows the overall learning-to-communicate framework. All the components are represented with NNs, and all the weights are trained end-to-end in a data-driven manner. Since the communication modules are shared across agents and the decoder output is updated in real-time using communication, we refer to the predicted state as the Common Operational Picture (COP). Future work can extend our data-driven COP framework to incorporate future actions, such as agent intent.

Since the communication is grounded in an interpretable state, human operators can use the learned black boxes to receive and interpret the communication between autonomous platforms. Similarly, human operators can encode their observations and transmit them using $\mathcal{C}_{out}$. In the rest of this paper, we do not consider human operators and assume the agents to be autonomous platforms.

The multi-agent system models a swarm performing a joint task, e.g., autonomous platforms in a C2 operation. The joint task is captured as an overall system reward function $\mathcal{R}: \mathcal{S} \times \mathfrak{A}^1 \times \ldots \times \mathfrak{A}^N \to [-1, +1]$. For example, the reward can capture mission success rate, Blue Force casualties, attrition rate, task completion, etc.

The agents are trained to maximize the reward in a centralized manner within the framework of Centralized Training Decentralized Execution (CTDE). The result of training is a decentralized policy $\pi^i: \mathcal{O}^i \times \mathbb{R}^{C^N} \to \mathfrak{A}^i$. During deployment, each agent executes the policy as a function of local observation and inter-agent communication.

The overall decision-making problem is defined as a Decentralized Partially Observable Markov Decision Process (DecPOMDP) $(N, P_0, \mathfrak{S}, \{\mathcal{O}^i\}, \{\mathfrak{A}^i\}, \mathcal{T}, \mathcal{R}, \{\mathfrak{P}^i\})$, $i = 1, \ldots, N$. $P_0$ defines the distribution over initial agent positions or "laydown" (friendly and enemy). $\mathcal{T}$ represents the transition function over time $s_{t+1} = \mathcal{T}(s_t, a_t^1, \ldots, a_t^N)$ and $s_0 \sim P_0$ that captures the evolution of state and the effect of the agent's actions.

We evaluate the robustness of the produced COPs and policies using a separate "test" DecPOMDP. In this paper, we focus on varying the capabilities of agents $\mathfrak{P}^i$ and laydown distribution $P_0$. The DecPOMDPs are incorporated into a simulator. In this paper we use the Starcraft-2 real-time strategy game [3] to simulate Command-and-Control (C2) scenarios.

## 2.1 BACKGROUND: QMIX FOR MULTI-AGENT LEARNING

QMIX [4] is a standard multi-agent Reinforcement Learning (MARL) algorithm within the CTDE framework. It extends Q-learning [5] to the multi-agent setting. QMIX learns a Q-function for each agent $Q: \mathcal{O}^i \times \mathfrak{A}^i \to \mathbb{R}$, mapping agent observation and action to a value. The Q-function defines an agent policy $\pi$ e.g., defined by the action with maximum Q-value. $Q(o^i, a^i)$ is the total reward of executing $a^i$ and then following the policy.

QMIX trains another centralized NN to model the Q-value of joint actions as a function of the Q-values of individual agent actions. In the simplest version, the centralized Q-function is a linear combination of agent Q-values. The coefficients of the linear combination are learned as a function of the global state $s_t$ (through a "hypernet" NN).

$$Q^{mix}(s, a^1, \ldots, a^N) = \sum_{i=1}^{N} \theta_i(s) Q(o^i, a^i)$$

In this model, the Q-function has only first-order terms, but higher-order terms can be included to capture correlations between agent Q-functions. The agent Q-functions, centralized Q-function and the hypernet are trained end-to-end with a temporal difference (TD) error:

$$\inf_{Q, \theta} \Big( Q^{mix}(s_t, a_t^1, \ldots, a_t^N) \\ - \big( r_t + Q^{mix}(s_{t+1}, a_{t+1}^1, \ldots, a_{t+1}^N) \big) \Big)^2 \quad (1)$$

where $r_t$ is the system reward at time $t$ for taking actions $a_t = (a_t^1, a_t^2, \ldots, a_t^N)$ in state $s_t$. We are significantly simplifying the exposition of QMIX [4] and deep Q-learning [6] and leaving out key details like exploration, target network, double Q-learning [7], etc. Note that the state is available during centralized training only and not available during deployment. Similarly, the centralized Q-function is used for training and not during deployment. QMIX does not use inter-agent communication.

## 3  COP MODEL AND TRAINING

In this section, we describe the neural network (NN) architecture and training for the communication modules $\mathcal{C}_{in}$ and $\mathcal{C}_{out}$, decoder $\mathcal{D}$, and the policy $\pi$. We motivate the design choices that will be investigated in Section 4. Note that our COP model can be optimized with any MARL algorithm such as QMIX.

We use the autoencoder (AE) concept as follows: an NN encoder transforms an agent observation to a latent vector $z$. Paired with the encoder is an NN decoder that reconstructs the observation given the latent vector $z$. An AE is trained end-to-end to minimize reconstruction error, e.g., mean squared error (MSE).

Following [17], instead of communicating "raw" observations, we can communicate the compact latent vector $z$ and ensure that the embedding vectors correspond to the agent observations. The embeddings are not easily interpretable by the adversary allowing for secure message transfer. However, there are two significant challenges in the application of AE to COP.

Suppose each agent encodes its own observation $o_t^i$ at the current time step and action $a_{t-1}^i$ of the previous time step and transmits the latent vector. Note that the observation and action are in the agents' frame of reference. On the receiver side at time $t + 1$, an agent (that can receive the message) can use an NN decoder $D$ to decode the sender's observation. However, the decoding will be in the senders' frame of reference.

Secondly, communication of agent observations alone is insufficient. Consider agents in a linear chain where each agent can only communicate with its neighbor. For the first agent's observation to reach the last agent, each agent must not only transmit its own (encoded) observation, but also (re-)transmit the message received from its neighbor. In general, the number of messages can grow exponentially (e.g., in tree-shaped connectivity structure). Rather, the agent must *integrate* the received observation (after decoding) and its own observation into a *local* COP. Then, only the latent vector corresponding to the local COP needs to be transmitted.

In this distributed setting, local integration of COP leads to predictions about agents outside any given agents' visual range. The local COP, or an aggregation of local COPs, should be interpreted as a density over the entire battlefield. Based on prior information, recency of observation and communication, the COP should have high density around likely agent positions. The spread or uncertainty should decrease with the frequency of observation of a given part of the battlefield by any agent.

We used the agent position attribute as an example to motivate the challenges in COP prediction and our distributed solution. In practice, our COP contains all information relevant to the state of an operation, e.g. agent armor, health, weapon status etc., each with an associated uncertainty (e.g., standard deviation). Our data-driven learning method, where ground truth is used to train each agent's COP model, is general to all attributes. We do not need to assume a sensor noise model or agent motion model in order to capture the uncertainty in each attribute represented in the COP.

### 3.1 Observation Encoding and Decoding

We apply a standard AE to the observation and action. Each agent encodes its own observation $o_t^i$ at the current time step and action $a_{t-1}^i$ of the previous time step using NN encoder $E_{obs}$, and transmits the embedding latent vector $z_t^i = E_{obs}(o_t^i, a_{t-1}^i)$. We use a $tanh$ activation on the encoder output so that each entry in the vector $z_t^i$ is bounded in the range $[-1, +1]$.

We consider two versions of the observation decoder:

- An explicit decoder NN $D_{obs}$ that maps $z_t^i$ to $(\widehat{o_t^i}, \widehat{a_{t-1}^i})$. The observation AE is trained to minimize

$$\inf_{E_{obs}, D_{obs}} \left\| o_t^i - \widehat{o_t^i} \right\|_2 + \left\| a_{t-1}^i - \widehat{a_{t-1}^i} \right\|_2 \quad (2)$$

- An implicit decoder that is part of a COP decoder, where reconstruction $(\widehat{o_t^i}, \widehat{a_{t-1}^i})$ is extracted from the COP and the known agent capabilities (e.g., FoV).

### 3.2 Integration of Communication

As described above, each agent must integrate received communications and produce an embedding vector corresponding to the COP. The embedding should be informative as an additional input to agent policies. The embedding space is shared across agents to enable a shared situational understanding among agents and human operators.

What type of NN architecture should be used to integrate communication? All received communications, from all agents can be potentially informative. Thus, we choose an attention-based architecture [8], which uses all-to-all connections, over an architecture based on convolutions that capture local connections. The attention-based architecture can learn to selectively attend to certain messages as a function of inputs, which is more powerful than fully connected architectures (like perceptron).

We use cross-attention [8] to process two sequences. Let $z_s^i$ be the integrated COP embedding vector. Let $\boldsymbol{c} = [z^1, \dots, z^N]$ and $\boldsymbol{c^s} = [z_s^1, \dots, z_s^N]$ be the embeddings received by the $i$th agent ($z^i$ as in Eq. 2). Then,

$$z_s^i \propto softmax(QK^T)V$$

where $Q = W_q \boldsymbol{c}$, $K = W_k \boldsymbol{c}$, $V = W_v \boldsymbol{c^s}$ where $W$ are learned matrices (in practice, multiple NN layers are stacked to form $z_s^i$). The incoming message processing module $\mathcal{C}_{in}$ (Section 2) is defined as $\mathcal{C}_{in}(\boldsymbol{c}, \boldsymbol{c^s}) \triangleq z_s^i$. In practice, several rounds of communication are used to propagate information and achieve consensus, by running the communication modules for multiple iterations so

$$z_s = \mathcal{C}_{in}\left(\mathcal{C}_{out}\left(\dots(\mathcal{C}_{in}(m))\right)\right).$$

Next, we track the temporal evolution of the COP using a recurrent NN (e.g., GRU [9]) that updates $h_s = GRU(z_s)$ using the history of $z_s$ values. It is this $h_s$ that we pass to the policy and the COP decoder.

We train a COP decoder NN $\mathcal{D}_{cop}$ with the ground truth egocentric state $s^i = f(s, i), s \in \mathcal{S}$, where $f$ is centering the state on the agent. For example, centering translates the positions of all units relative to the position of agent $i$.

$$\inf_{\mathcal{D}_{cop}} \frac{1}{N} \sum_{i=1}^{N} \left\| s_t^i - \mathcal{D}_{cop}(h_s) \right\|_2 \quad (3)$$
$$h_s = GRU(z_s)$$
$$z_s = \mathcal{C}_{in}(\boldsymbol{m_t})$$

where $s_t$ is ground truth state provided by the simulator, $\boldsymbol{m_t} = [\mathcal{C}_{out}(o_{t-1}^1, a_{t-1}^1), \dots, \mathcal{C}_{out}(o_{t-1}^N, a_{t-1}^N)]$ is the concatenation of the $C$-dimensional messages received from the agents from the previous time step,

$$\mathcal{C}_{out}(o^i, a^i) \triangleq [z, z_s] = [E_{obs}(o, a), \mathcal{C}_{in}(c, c_s)]$$

applied to the previous time step observation and action.

Note that the agents transmit both $z$ and $z_s$ to mitigate two different sources of error: $z$ controls the error in reconstruction of agent's observation (within field-of-view) and $z_s$ controls the error in prediction of unobserved enemies (outside the field-of-view).

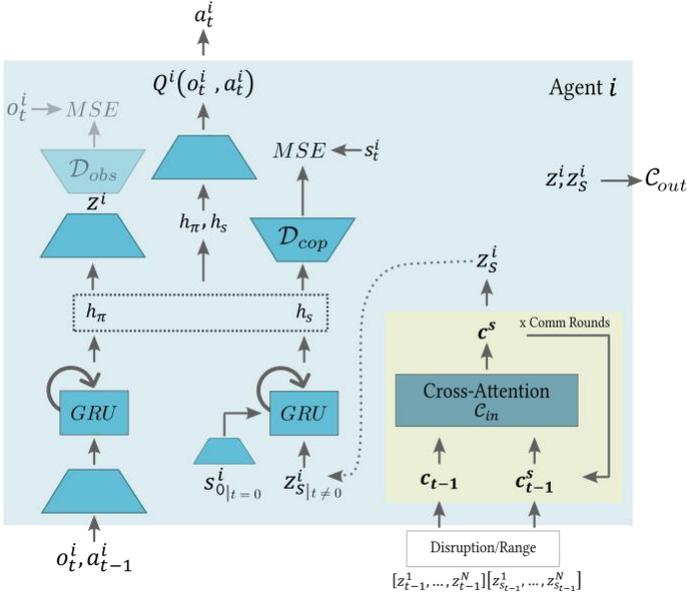

*Figure 4: Overview of the agent architecture. The output is the agent action $a_t^i$ and agent communication $(z, z_s)$. The observation decoder $D_{obs}$ is optional.*

### 3.3 Incorporating Initial State

In some scenarios, information about the initial state is known (e.g. initial enemy positions gathered from ISR). Incorporating a known initial state can significantly increase the accuracy of predicted COPs because it provides ground truth positions especially for enemy units outside the field-of-view of all friendly units. The COP model can learn to *track* the changes in the state by decoding the friendly actions and projecting the effect on the state. By using different initial states, we can increase the diversity of the training set.

When an initial state $s_0$ is provided (it is optional), we use it to initialize the hidden state of the GRU (Figure 4). The initial state is projected to the dimensions of the GRU hidden state using one linear layer. $W_{init}$ is trained end-to-end alongside the other parts of the model.

$$h_0 = W_{init}.s_0$$

### 3.4 Hallucination

Note that the COP prediction is over a generic set of attributes including position $x$ and health attributes $h$, etc., over all the units. The loss function in Eq. (2) and Eq. (3) gives uniform weightage to all attributes. We add a term to the training objective Eq. (4) to explicitly penalize hallucinations of agents that are present in the COP, but, in fact, not present in the simulation due to zero health.

$$\lambda_h(1-H).\left(H - H_{cop}\right)^2 \qquad (4)$$

where $H$ is the ground truth health (from the state) and

$H_{cop}$ is the predicted health predicted by the COP model. The term is added for all agents and simulation steps.

### 3.5 Integrating COP with Policy Learning

As mentioned before, the COP $h_s$ is provided as an additional input to the policy via the GRU. Now we describe the policy architecture. Following QMIX (Section 2.1), the policy first processes the observation and action using linear layers. Then, we use a recurrent NN (e.g., GRU) to capture the observation and action history. The hidden state of the policy GRU $h_\pi$ and the hidden state of the COP GRU are concatenated. The agent Q-value function takes the concatenated hidden state (denoted $[h_s, h_\pi]$) and uses linear layers to map to action values.

Figure 4 shows the NN architecture for COP and policy. The overall training objective is a combination of policy and COP training objectives from Eq. (1), (2), (3), (4):

$$\inf_{Q, E_{obs}, D_{cop}, D_{obs}} \left( Q^{mix}(s_t, a_t^1, \dots, a_t^N) \right.$$
$$\left. - \left( r_t + Q^{mix}(s_{t+1}, a_{t+1}^1, \dots, a_{t+1}^N) \right) \right)^2$$
$$+ \frac{1}{N} \sum_{i=1}^{N} \left\| s_t^i - D_{cop}(h_s) \right\|_2$$
$$+ \left\| o_t^i - \hat{o_t^i} \right\|_2 + \left\| a_{t-1}^i - \widehat{a_{t-1}^i} \right\|_2$$
$$+ \lambda_h(1-H).\left( H - H_{cop} \right)^2 \qquad (5)$$

## 4 Experiments

In this section, we evaluate the data-driven approach for COP prediction and its impact on multi-agent policy learning. While ML-driven methods such as ours can be highly precise, they can be fragile and may deviate from the training scenarios or simulations. Therefore, we evaluate the resilience of the COP and the policy on separate test scenarios [12].

### 4.1 Observation and Action Spaces

As mentioned earlier, we use the Starcraft-2 (SC2) [3] game environment to study Command-and-Control (C2) scenarios. We use the Starcraft Multi-Agent Challenge (SMAC [1] and SMACv2 [10]) that instruments the simulator and provides a multi-agent interface. We use the pyMARL software library [1] and build on the algorithm implementations therein. To study heterogenous agents, we extend pyMARL and add agent-specific capabilities:

- Sight range: circular field-of-view in pixels.

- GPS: 2D (X,Y). Set to zeros when GPS is jammed.

- Shoot range: in pixels.

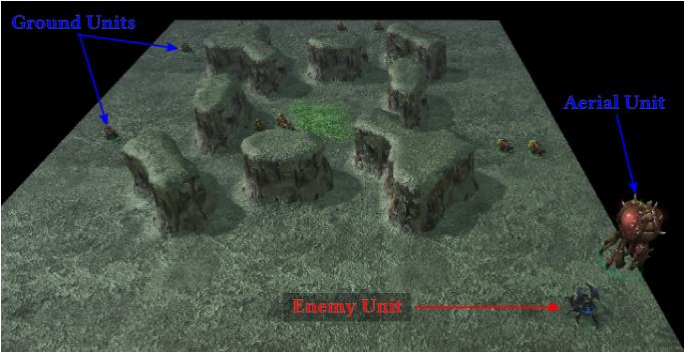

*Figure 5: Air-Ground Recon ("1O10B-vs-1R") [11].*

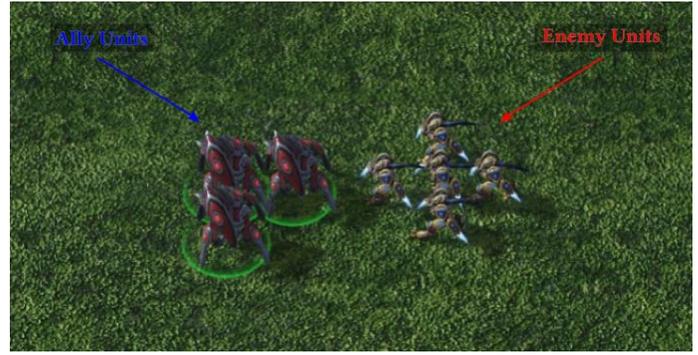

*Figure 6: Withdrawing Attack ("3S-vs-5Z") [1].*

In this paper, we work with fixed-size observation and state spaces. We assume that the maximum number of agents is known beforehand. When the number of agents is lesser than the maximum (due to limited field-of-view, or dead agents), the corresponding entries in the observation and state vectors are filled with zeros. Geographical features are not observed. A given agent observes non-zero entries for other agents within its field-of-view. The real-valued observation vector contains:

- Distance (in pixels) from the observing agent to the observed agent.

- Relative coordinates of the observed agent to the observing agent (egocentric, 2D).

- Health, shield, and type of the observed agent.

- Whether observed agent is within shoot range.

- Observing agents own health, shield, unit type.

Each agent can execute movement actions (in four cardinal directions N, S, E, W by a fixed number of pixels) and attack actions (one action per enemy agent within shoot range). The state vector contains the following information for each agent:

- Absolute position (in global coordinates).

- Health, shield, weapon status, unit type.

Among these, the coordinates of agents require transformation between egocentric and global frames.

## 4.2 METRICS

We evaluate all these metrics on training and test scenarios. We evaluate if our method produces a COP that matches the ground truth in terms of the Mean Squared Error (MSE). We compute the MSE for (1) the predicted COP and the ground truth state (from the simulator), and (2) the predicted field-of-view of all agents and the ground truth field-of-view. For each time step, the MSE is averaged over the agents and episode length, and the goal is to achieve a normalized MSE < 5% over the friendly and enemy features (location, health, etc.).

We define a key metric for COP prediction called hallucination. Hallucination refers to the prediction that an agent has non-zero health in the COP when the agent has zero health (dead agent) in the ground truth (simulation). Hallucination is calculated as the average error in the predicted health over dead agents.

Finally, we evaluate the success of the policy w.r.t. a clearly defined win condition. We report the average win rate (over 5000 episodes) for training and test scenarios.

## 4.3 SCENARIOS

***Air-Ground Recon***: This scenario tests the ability of a friendly aerial agent to track the movements of an enemy ground unit and communicate its position in a manner that friendly ground forces can decode. Upon decoding, friendly forces must move to attack the enemy before the time expires. In the Starcraft-2 (SC2) simulation, we used the scenario "*1O10B-vs-1R*" (shown in Figure 5 introduced in [11]). It contains one friendly aerial unit, ten friendly ground units, and one enemy unit. The enemy must be attacked by all ten friendly units to be defeated. The win condition is to kill the enemy unit before the timer expires.

***Withdrawing Attack***: In this scenario, the enemy force outnumbers the friendly force five to three. However, the friendly agents have a speed advantage while in retreat. The friendly units must jointly attack and withdraw in a coordinated fashion to evenly distribute the damage from enemy attacks across the friendly units by performing a "kiting" micromanagement strategy. Precise coordination in this scenario requires precise COPs. In SC2, we used the "3S-vs-5Z" scenario [1]. The win condition is to kill all enemy agents before the timer expires.

***TigerClaw***: The *TigerClaw* melee map [13] is a high-level recreation of the TigerClaw combat scenario (Figure 7) developed using the StarCraft-2 map editor. The scenario ("*TC_5B-vs-6R*") was developed by Army subject-matter experts (SMEs) at the Captain's Career Course, Fort

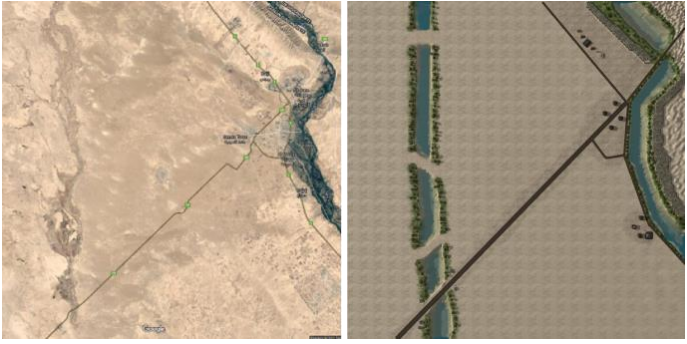

*Figure 7: TigerClaw: (Left) Original geographical map. (Right) Corresponding designed map in StarCraft-2.*

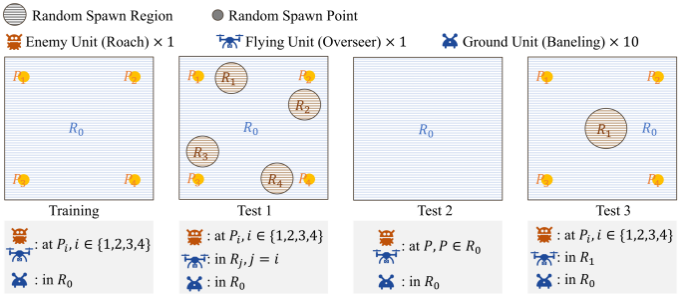

*Figure 8: OOD maps for Air-Ground Recon (1O1B-vs-1R).*

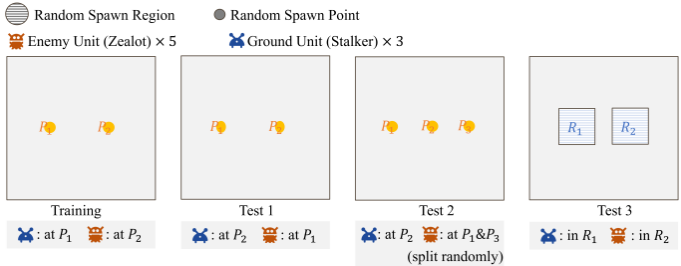

*Figure 9: OOD maps for Withdrawing Attack (3S-vs-5Z).*

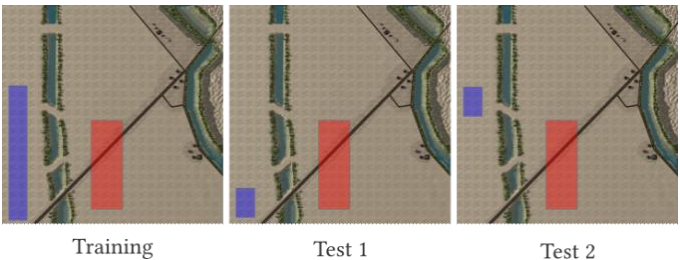

*Figure 10: Tigerclaw OOD maps. Blue and red regions indicate the spawning areas for the blue and the red forces, respectively.*

Moore, Georgia. The Blue Force is an Armored Task Force (TF), which consists of combat armor with M1A2 Abrams, mechanized infantry with Bradley Fighting Vehicles (BFV), mortar, armored recon cavalry with BFV, and combat aviation. The Red Force is a Battalion Tactical Group (BTG) with attached artillery battery and consists of mechanized infantry with BMP, mobile artillery, armored recon cavalry, combat aviation, anti-armor with anti-tank guided missiles (ATGM), and combat infantry. As seen in Figure 7, the terrain is challenging in this scenario because there are only four viable wadi crossing points. The win condition is to kill 80% of the red agents. The red force uses fixed behavior rules in all our scenarios.

## 4.4 TEST SCENARIOS

We evaluate our trained policies on modified laydowns (change in the initial positions of friendly and enemy) in all scenarios. These Out-of-Distribution (OOD) maps helps understand the generalization capability of our method.

OOD maps for Air-Ground Recon ("1O1B-vs-1R") and Withdrawing Attack ("*3S-vs-5Z*") are showing in Figure 8 and Figure 9 (for more details refer to [12]). In *TigerClaw*, we change the blue force spawning region from defending all crossings to either the south wadi or further north, as shown in Figure 10. In training and testing, blue force is randomly spawned in the corresponding blue region.

## 4.5 COMPARISON METHODS AND BASELINES

Our method of prediction of COPs is general and can be learned with a fixed policy, or jointly learned with any MARL method. In this paper, we integrated COP learning into the QMIX [4] MARL method. Future work can explore integration with more recent MARL methods. Note that QMIX does not use inter-agent communication. We built two strong baselines on top of QMIX to compare against.

- *QMIX w/ $s_0$*: A version of QMIX leveraging the initial state knowledge. This is a strong baseline because it alleviates the issue of partial observability. In this baseline, each agent takes an additional input, i.e., the initial state vector, in addition to the agent observation input.

- *QMIX w/ Cross Attention*: A baseline that incorporates inter-agent communication into QMIX. Each agent receives an additional input, a message containing the *raw* observations of all the agents. The agent architecture is modified to process these observations using Cross Attention.

We compare to recent multi-agent RL methods that learn the inter-agent communication function: (1) *MASIA* [14] predicts the state from other agents' observations within a QMIX method, (2) *NDQ* [11] and (3) *TarMAC* [15] are prior work on MARL where learned communication is not grounded on state prediction.

## 4.6 HYPERPARAMETERS

Training is performed end-to-end using the training objective in Eq. (4). We perform training in simulation for prediction separately to evaluate the effect of denied GPS. Note that XY prediction requires a change in the frame of reference. In all scenarios, the COP model learns to predict XY positions (< 0.05, bottom panel) more accurately than the health attribute (~0.1, top panel).

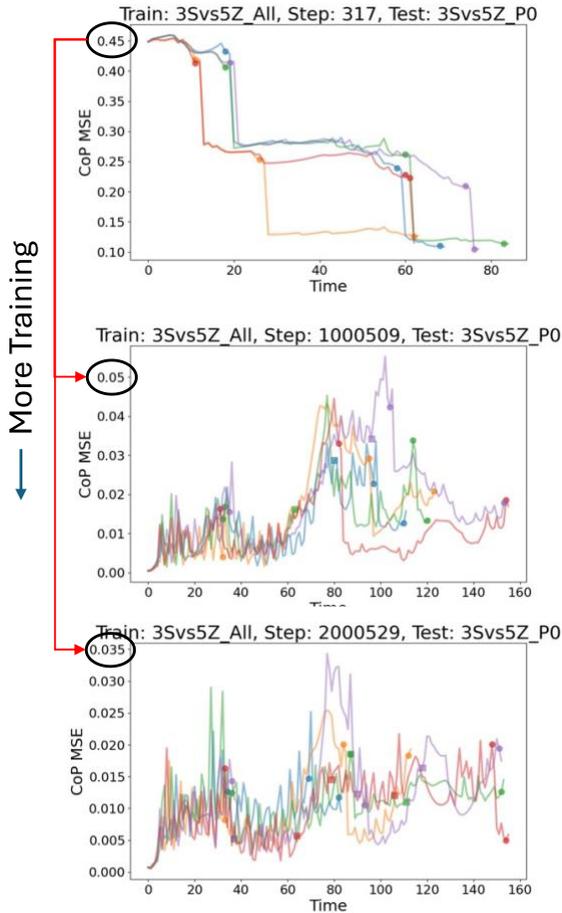

*Figure 11: COP MSE over episode (color-coded by different playout episodes) for different stages of training. Dots mark the timestep that an agent was killed.*

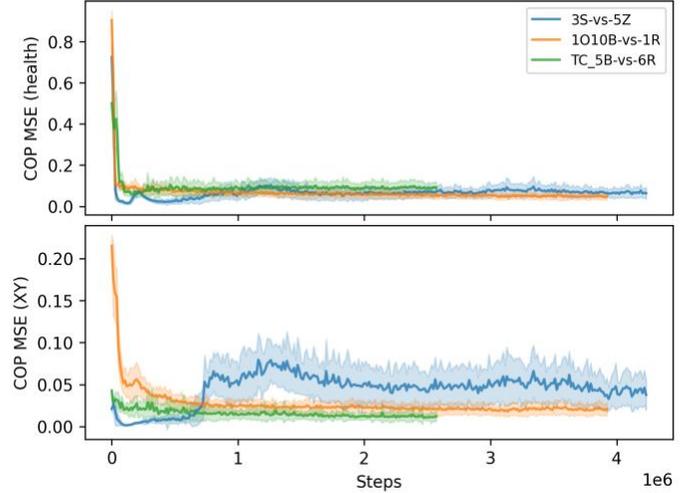

Figure 12: COP MSE for health and XY prediction across all scenarios.

In *3S-vs-5Z*, while the MSE is low, there is a noticeable increase in COP error over training (blue line), but not in other scenarios where the COP error decreases monotonically. The different dynamics reflect the differences between the scenarios. To succeed in the *3S-vs-5Z* scenario, the policy needs to learn to evade, withdraw, and attack. The training seems to tradeoff COP errors and focus on policy learning. In the other two scenarios (green and orange lines), accurate prediction of enemy agents' locations is required to win. Hence, we see stable and accurate COP XY throughout training.

### 4.7.2  Human Interpretability

A key feature of our method is that the COPs are human-interpretable. Figure 13 shows a visualization of the state and corresponding predicted COP for all three scenarios.

In the example from the *3S-vs-5Z* scenario (first row, left panel), three enemy agents are engaging one friendly unit on the far east side of the map. We see that the friendly can successfully communicate the positions of enemies (first row, right panel) to friendly agents on the far west side of the map. In the second example (middle row) from the *1O10B-vs-1R* scenario, we see that the position of the enemy (unit marked "EO" in the top left of the map) is accurately communicated by the aerial unit and represented with low uncertainty, meaning that the model can leverage heterogeneous agent capabilities. In the third example (third row) from the TigerClaw scenario, we observe that the model can predict friendly positions

---

20 million steps. We use the Adam optimizer with learning rate of 1e-3. Gradients are clipped at a norm of 20. Policy updates uses $\lambda_{td} = 0.3$ and $\gamma = 0.99$. We set $C = 32$, (communication dimension) and $h_s = 64$ (hidden dimension) with four cross-attention heads. By default, hallucination penalty $\lambda_h = 3$. For exploration, we use $\epsilon$-greedy exploration, annealing $\epsilon$ from 1 to 0.05 over a scenario-dependent number of steps, typically $10^5$ steps, except in *TigerClaw*, we use a schedule of $5x10^5$ steps.

### 4.7  Results

#### 4.7.1  Convergence of COP under Denied GPS

First, we evaluate the model without GPS capability for any agent. We track the MSE of the COP against the ground truth state. As seen in Figure 11, in the "3s-vs-5z" scenario, the MSE is initially high (top panel), and with further training, the COP converges to low MSE (bottom panel), showing that the COP model produces highly accurate COPs.

Figure 12 shows the convergence of COP in all three scenarios. We show the COP MSE for health and XY

State    Decoded COP

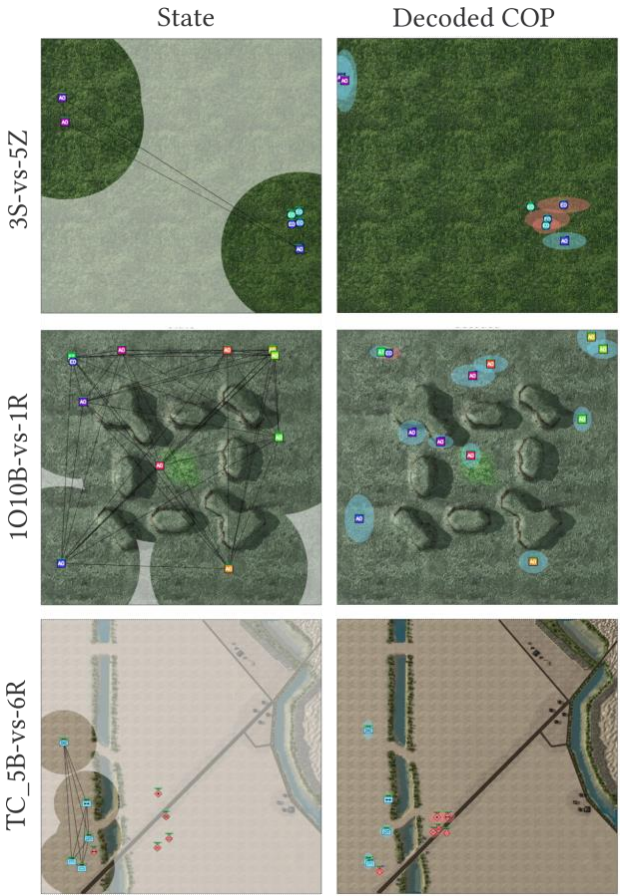

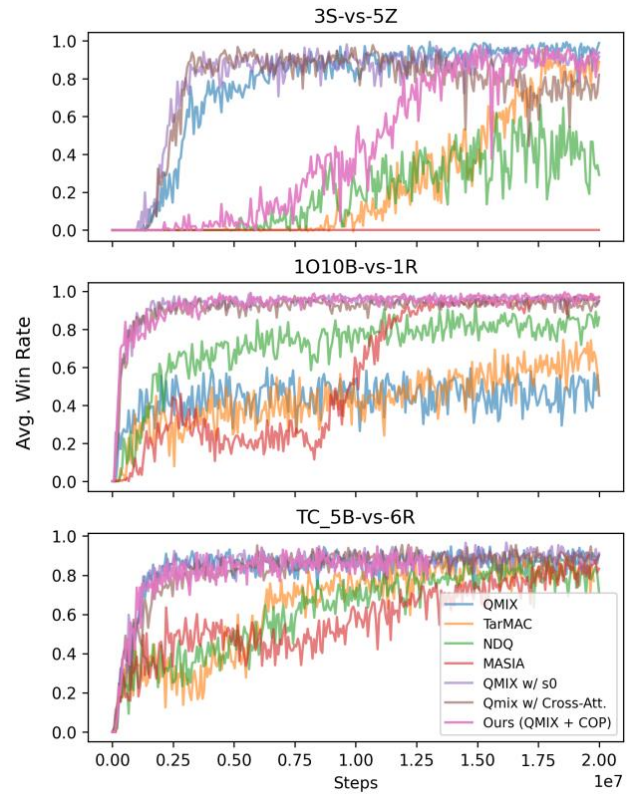

*Figure 14: Average Win Rate across training steps for Our method and other baselines*

*Figure 13: (Left) State. Gray regions represent fog; clear regions represent field-of-view. Lines connecting agents represent communication. (Right) Human interpretable predicted COP. Ellipses represent uncertainty and consensus: standard deviation in predicted agent position over the COPs from all agents.*

accurately. Still, the enemy positions are not accurate beyond the visual range.

To visualize the uncertainty in the COP, we averaged the predicted XY positions across all agents and showed axis-aligned ellipses with major and minor axes equal to the standard deviation in X and Y coordinates. A small ellipse means low uncertainty and high consensus among agents.

### 4.7.3    COP Improves Convergence Rate of Policy

We trained the agent policies jointly with the COP models. The training results are shown in Figure 14. As expected, QMIX is unable to solve the Air-Ground Recon problem (second row) as it does not use communication. The scenario timer expires before the enemy position is discovered. The baseline *QMIX w/ Cross Attention* directly uses the raw observations of all agents, it does not suffer from the partial observability issue. Similarly, *QMIX w/ $s_0$* has access to the initial state (including enemy position) which is highly informative in the *1O10B-vs-1R* scenario.

This explains the fast convergence of both the baselines to winning policies.

In comparison with methods that learn to communicate, our COP-based method outperforms *TarMAC* [15] (orange line), *NDQ* [11] (green line), and *MASIA* [14] (red line), in all three scenarios. In two out of three tested scenarios, our method is more than an order of magnitude faster to converge. Faster convergence or lower sample complexity reduces the time and effort required to produce such multi-agent policies for autonomous platforms. The results show that our method provides an inductive bias based on situational awareness that facilitates learning. In the next sections, policies are evaluated for robustness and resilience to variations between training and testing.

### 4.7.4    Resilience to Fog

In this section, we test the trained policies in scenarios where the agents' sight range is reduced. Table 1 shows the results where the sight range of agents' is reduced by 33%-66% of the training sight range. For example, in the *TigerClaw* scenario, the agents use a sight range of 30 pixels for training, i.e., they can observe enemies up to 30 pixels away.

When we reduce the sight range, we notice that the performance of *QMIX* and *QMIX w/ $s_0$* (these do not use communication) drops significantly (0.88/0.91 win rate down to 0.43/0.53 win rate). In comparison, our method is significantly more robust (0.9 at 30 pixels to 0.64 at 10 pixels), meaning the intelligent inter-agent communication mitigates the degradation in the visual range. We also observe the performance of *QMIX w/ Cross Attention* at visual range of 20 pixels is lower than our method (0.74 vs 0.81 for our method). This difference is significantly higher in the *3S-vs-5Z* scenario (0.11 vs 0.94 for our method). We see that our method is the most robust in all three scenarios with variations in sight range.

Table 1: Resilience of Policies trained in conjunction with COP to variation in the sight range of agents.

| Method | Sight Range | | |
|---|---|---|---|
| | Train | Test | |
| **TC_5B-vs-6R** | 30 pixels | 20 px | 10 px |
| QMIX | 0.88 | 0.71 | 0.43 |
| QMIX w/ $s_0$ | 0.91 | 0.76 | 0.53 |
| QMIX w/ Cross Attn. | 0.88 | 0.74 | 0.65 |
| QMIX+COP (Ours) | 0.9 | **0.81** | **0.64** |
| **3Svs5Z** | 9 pixels | 6 px | 3 px |
| QMIX | 1.0 | 0.79 | 0.0 |
| QMIX w/ $s_0$ | 1.0 | 0.57 | 0.0 |
| QMIX w/ Cross Attn. | 0.99 | 0.11 | 0.0 |
| QMIX+COP (Ours) | 1.0 | **0.94** | 0.0 |
| **1O10B-vs-1R** | 9 pixels | 6 px | 3 px |
| QMIX | 0.47 | 0.47 | 0.43 |
| QMIX w/ $s_0$ | 0.99 | 0.99 | **0.98** |
| QMIX w/ Cross Attn. | 0.98 | 0.96 | 0.91 |
| QMIX+COP (Ours) | **0.99** | **0.98** | 0.92 |

### 4.7.5 Resilience to Denied GPS

In this experiment, agents have GPS enabled during training and are not denied. We compare the performance to an upper bound of retraining without GPS. We test two scenarios: partial denial of GPS and total denial of GPS. In the Air-Ground Recon scenario (*1O10B-*

Table 2: Resilience of Policies trained with GPS, to testing with GPS denial. Performance of policies trained with total GPS denial is shown in parenthesis as an upper bound.

| Test GPS Config. | | | |
|---|---|---|---|
| Scenario | All | Partial | Total |
| 1O10B-vs-1R | 0.97 | 0.68 | 0.68 (0.94) |
| 3s-vs-5z | 0.93 | 0.83 | 0.7 (0.7) |
| TC_5B-vs-6R | 0.87 | 0.85 | 0.73 (0.88) |

*vs-1R*), we deny GPS for all ground units, and only the aerial unit has GPS. In the Withdrawing Attack scenario (3S-vs-5Z), we deny GPS for all agents except one selected at random. In the TigerClaw scenario, we deny GPS for all agents except the helicopter and scout platoons. We report the win rate degradation in Table 2.

In the *Air-Ground Recon* scenario, we see a significant degradation (0.97 vs 0.68 win rate) compared to the upper bound win rate of 0.94. In the *Withdrawing Attack* scenario, there is degradation in win rate (0.93 down to 0.83 and 0.7). However, the performance matches the upper bound. That is, our method achieves the same performance without retraining the policies without GPS. Overall, in all three scenarios we see that our method can mitigate the degradation even if one of the agents has GPS (e.g., win rate 0.85 vs 0.73 in *TigerClaw*).

### 4.7.6 Resilience to Change in Enemy Laydown

A potential vulnerability of MARL trained policies is when the initial state distribution is different to the training configuration. Specifically, a change in the initial enemy positions, aka laydown, can cause degradation in win rate. We tested the trained policies on the OOD scenarios from Section 4.4. The results are shown in Table 3.

Among methods that learn to communicate, our method based on COP outperforms prior methods *TarMAC, NDQ,* and *MASIA*. In the *TigerClaw* scenario, our method retains the win rate on test laydowns as well. In the *1O10B-vs-1R* scenario (Test 3), our method is almost twice as effective (0.78 vs 0.41 Win-Rate). **The average win rate over scenarios and laydowns for our method is 0.837 vs the second-best 0.645 TarMAC.**

## 4.8 MODEL ABLATIONS RESULTS

### 4.8.1 Hallucination

As mentioned in Section 3.4, hallucination of agents is an issue in our COP model. Figure 15 shows an example of hallucination in the *TigerClaw* scenario. Four red agents are reported in the COP, including the artillery unit. Out of four, two agents are dead in the simulation, while another red agent (this is alive) is not captured in the COP.

*Table 3: Average win rate (5000 Episodes) on OOD laydowns for different scenarios. Boldface numbers represent best performance among methods that learn to communicate (excluding baselines in the first group of rows).*

| Scenario | 3S-vs-5Z | | | | 1O10B-vs-1R | | | | TC_5B-vs-6R | | |
|---|---|---|---|---|---|---|---|---|---|---|---|
| **Laydown** | Train | Test1 | Test2 | Test3 | Train | Test1 | Test2 | Test3 | Train | Test1 | Test2 |
| QMIX [4] | 1.0 | 0.87 | 0.75 | 0.91 | 0.47 | 0.45 | 0.29 | 0.4 | 0.88 | 0.84 | 0.73 |
| QMIX w/ s0 | 1.0 | 0.11 | 0.74 | 0.71 | 0.99 | 0.98 | 0.74 | 0.85 | 0.91 | 0.89 | 0.79 |
| QMIX w/ Cross-Att. | 0.99 | 0.85 | 0.94 | 0.8 | 0.98 | 0.96 | 0.56 | 0.25 | 0.88 | 0.89 | 0.71 |
| TarMAC [15] | 0.92 | 0.62 | **0.93** | 0.46 | 0.62 | 0.6 | 0.39 | 0.2 | 0.82 | 0.76 | 0.77 |
| NDQ [11] | 0.64 | 0.2 | 0.4 | 0.39 | 0.89 | 0.85 | 0.35 | 0.34 | 0.76 | 0.77 | 0.72 |
| MASIA [14] | 0 | 0 | 0 | 0 | 0.95 | 0.87 | 0.47 | 0.41 | 0.71 | 0.73 | 0.71 |
| **QMIX + COP (Ours)** | **1.0** | **0.78** | 0.54 | **0.77** | **0.99** | **0.99** | **0.68** | **0.78** | **0.91** | **0.91** | **0.86** |

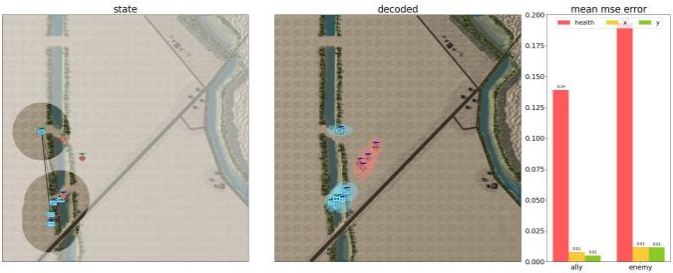

*Figure 15: An example COP showing the problem of hallucination, i.e. units are still tracked for a few steps after they are dead in simulation.*

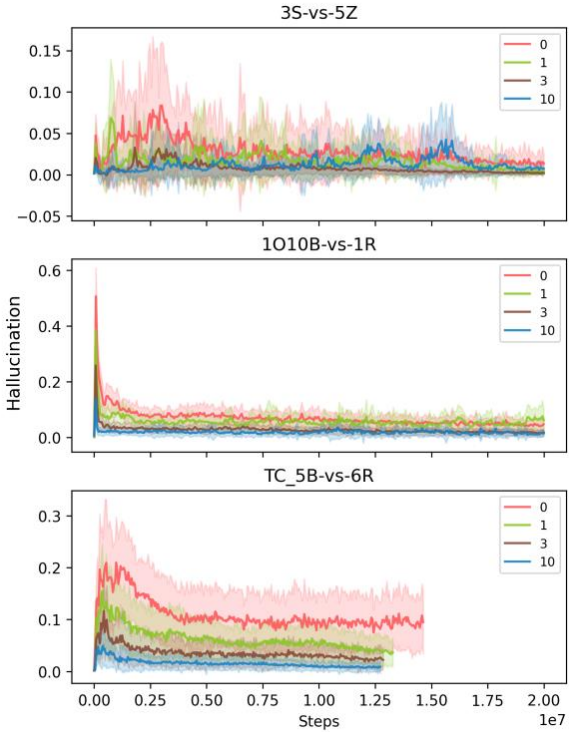

*Figure 16: Effect of hallucination penalty on hallucination over training steps for different scenarios*

*Table 4: Effect of three rounds of communication vs one round of communication on the average win rate.*

| Win Rate | Scenario | | |
|---|---|---|---|
| Rounds | 1O10B-vs-1R | 3s-vs-5z | TC_5B-vs-6R |
| 1 | 0.33 | 0.98 | 0.85 |
| 3 | 0.97 | 0.93 | 0.87 |

In general, we observed from the rollouts that dead agents are represented in the COP for a few time steps before they are reflected as dead. Even though this issue is reduced over training, there are still hallucinations observed in the trained model. Therefore, we introduced an explicit hallucination penalty as described in Section 3.4. We found that the increased weightage for health predictions significantly reduces the hallucinations as shown in Figure 16. We observed that the hallucination penalty $\lambda_h = 3$ significantly reduces the hallucination without reducing the win rate.

#### 4.8.2   Number of rounds of communication

We use multiple communication rounds per simulation step to refine the COP further and achieve a consensus among agents. Table 4 shows the average win rate comparing one vs three rounds of communication. The table shows that three rounds of communication achieve a higher win rate, especially in the 1O10B-vs-1R scenario.

## 5   DISCUSSION

The experiments showed resilience to novel scenarios:

GPS denial, communication denial (fog), and unknown enemy laydown. The experimental results show that the method produces precise COPs and highly resilient policies. We identified the issue of hallucination and introduced a training regularizer to control it.

Currently, most COP formation is performed manually in such challenging scenarios by collating communications at a C2 node. This process is too slow to produce the COP, a key data product for decision-making processes and autonomous policies that run on the platforms. The manual process is not scalable with the number of platforms and the amount of data to be processed. The paper addresses this barrier to future C2.

As shown in our experiments, this work also significantly advances multi-agent reinforcement learning (MARL). Existing MARL methods do not work well in our challenging scenarios. Among MARL methods that learn to communicate, existing methods suffer from a lack of grounding: it is unknowable what the swarm is communicating, whether it accurately captures an evolving scenario, and how a human operator can be brought into the loop. This can produce unintended or undesired behaviors in new scenarios.

In terms of future directions, multi-task training is a straightforward extension to make the COPs and policies even more resilient. We can randomly vary the scenario, sight range, communication range, laydown, capabilities, etc., during training. Future work should explore sparse communication (sparse in time). We did not vary the bandwidth requirement of communications in this paper and future work must explore this as well.

Multi-agent exploration of an unknown dynamic battlefield (e.g., ISR) is challenging as the number of agents grows. A promising direction for future research is to use our COPs for exploration. The COP can help identify areas of high uncertainty, and methods similar to active sensing can be derived to explore such regions of the battlefield in a coordinated manner.

A key unsolved modelling challenge is to support a variable number of agents. Future work can consider sequence-based models to allow messages of arbitrary length without dependence on the number of agents. In fact, the success of the cross-attention mechanism used in this paper indicates the likelihood of success of such models for COP formation.

Another key challenge is the comprehensive evaluation of COPs and policies. So far, we have manually explored the space of enemy laydowns to come up with challenging scenarios. Future work can explore scenario co-design methods (e.g., [16]) where challenging scenarios are created within the training loop.

In summary, this paper presents a data-driven method for common operational picture (COP) formation in a multi-agent system. The method works with general perceptions from heterogenous platforms in a GPS-denied environment. The COP is general, including but not limited to unit positions (vs. methods that only estimate positions, such as dead reckoning). The COP formation is fully autonomous, real-time, and human-interpretable.

### ACKNOWLEDGEMENT

This material is based upon work supported by the Army Research Laboratory (ARL) under the Army AI Innovation Institute (A2I2) program Contract No. W911NF-22-2-0137.

Robotics at CoRL 2023, 2023.